\documentclass[twocolumn,prl,aps,superscriptaddress]{revtex4}
\usepackage{epsf}
\usepackage{graphicx}
\usepackage{epstopdf}
\usepackage{booktabs}
\usepackage{multirow}
\usepackage{amsmath}
\usepackage{array}
\usepackage{float}
\usepackage{lipsum,color}
\begin{document}

\title{Non-deterministic self-assembly of two tile types on a lattice}

\author{S. Tesoro}
\affiliation{Theory of Condensed Matter, Cavendish Laboratory, University of Cambridge, CB3 0HE Cambridge, UK}
\author{S. E. Ahnert}
\affiliation{Theory of Condensed Matter, Cavendish Laboratory, University of Cambridge, CB3 0HE Cambridge, UK}

\begin{abstract}
Self-assembly is ubiquitous in nature, particularly in biology, where it underlies the formation of protein quaternary structure and protein aggregation. Quaternary structure assembles deterministically and performs a wide range of important functions in the cell, whereas protein aggregation is the hallmark of a number of diseases and represents a non-deterministic self-assembly process. Here we build on previous work on a lattice model of deterministic self-assembly to investigate non-deterministic self-assembly of single lattice tiles and mixtures of two tiles at varying relative concentrations. Despite limiting the simplicity of the model to two interface types, which results in 13 topologically distinct single tiles and 106 topologically distinct sets of two tiles, we observe a wide variety of concentration-dependent behaviours. Several two-tile sets display critical behaviours in form of a sharp transition from bound to unbound structures as the relative concentration of one tile to another increases. Other sets exhibit gradual monotonic changes in structural density, or non-monotonic changes, while again others show no concentration dependence at all. We catalogue this extensive range of behaviours and present a model that provides a reasonably good estimate of the critical concentrations for a subset of the critical transitions. In addition we show that the structures resulting from these tile sets are fractal, with one of two different fractal dimensions.
\end{abstract}

\maketitle

\section{Introduction}
Self-assembly is a widespread phenomenon that is observed across many biological, chemical, and physical systems. Examples include DNA \cite{winfree1998design,mao2000logical,chworos2004building,goodman2005rapid,rothemund2006folding,fujibayashi2007toward,ke2012three}, protein quaternary structure \cite{Levy:2006ez,villar2009self}, protein aggregation \cite{Cohen:2013fd}, viruses \cite{zlotnick1994build}, micelles \cite{israelachvili1994self}, and thin films \cite{krausch2002nanostructured}, among others. In biology it is of particular importance as protein self-assembly gives rise to an enormous variety of molecular machinery in every living cell. A simple lattice model of self-assembly was introduced in the recent literature and has found application in the study of structural complexity \cite{Modularity} as well as the biological evolution of self-assembly \cite{Evo} and the more general study of genotype-phenotype maps \cite{green}. The primary focus of this work has so far been deterministic self-assembly, meaning tile sets that produce the same final structure, even in a stochastic assembly process. Non-deterministic self-assembly however is of also of primary importance in biology, particularly in the context of disease, as uncontrolled protein aggregation underlies sickle-cell anaemia \cite{Bunn:1997de} as well as the formation of amyloid fibrils, which can cause a wide variety of different pathologies \cite{Cohen:2013fd}. Here we exhaustively study, for the same lattice model, all topologically distinct single tiles and two-tile sets with up to two different interfaces, considering both symmetric and asymmetric interactions. There are 13 such single tiles and 106 tile sets. Many of these assemblies are non-deterministic, meaning that a stochastic assembly process for a given tile set results in different structures every time the assembly is initiated from a single seed tile. We find that, despite the limited number of interactions, the non-deterministic two-tile sets display a wide variety of different assembly behaviours, and that in many cases the final structure exhibits a strong dependence on the relative concentrations of the two tiles. While the lattice model of self-assembly is vastly simpler than the biological self-assembly of proteins, it can nevertheless offer important insights. For example the haemoglobin complex, which consists of two copies of two different proteins, is a bound structure that can be represented using a two-tile set in the lattice model \cite{green}. The sickle-cell mutant of the complex can be represented by mutating one of the tiles, adding an additional interface. This leads to an unbound structure \cite{green,Bunn:1997de}. Similarly, amyloid fibrils, which form when proteins misfold and bind to other proteins in an uncontrolled fashion, can be modelled by starting with a single tile that forms a structure through deterministic self-assembly and then introducing a second tile with an alternative configuration of interactions, representing the mutated version of the same protein. Strong concentration dependence has been observed in such aggregation processes \cite{Cohen:2013fd}, which is also what we observe for a number of two-tile sets in our highly simplified lattice model. The model, while abstract, is therefore nevertheless capable of reproducing self-assembly behaviours that are analogous to non-deterministic assembly phenomena in biology.

\section{The self-assembly model}\label{rules}
\begin{figure}[]
 \centering
 \includegraphics[width=0.48\textwidth]{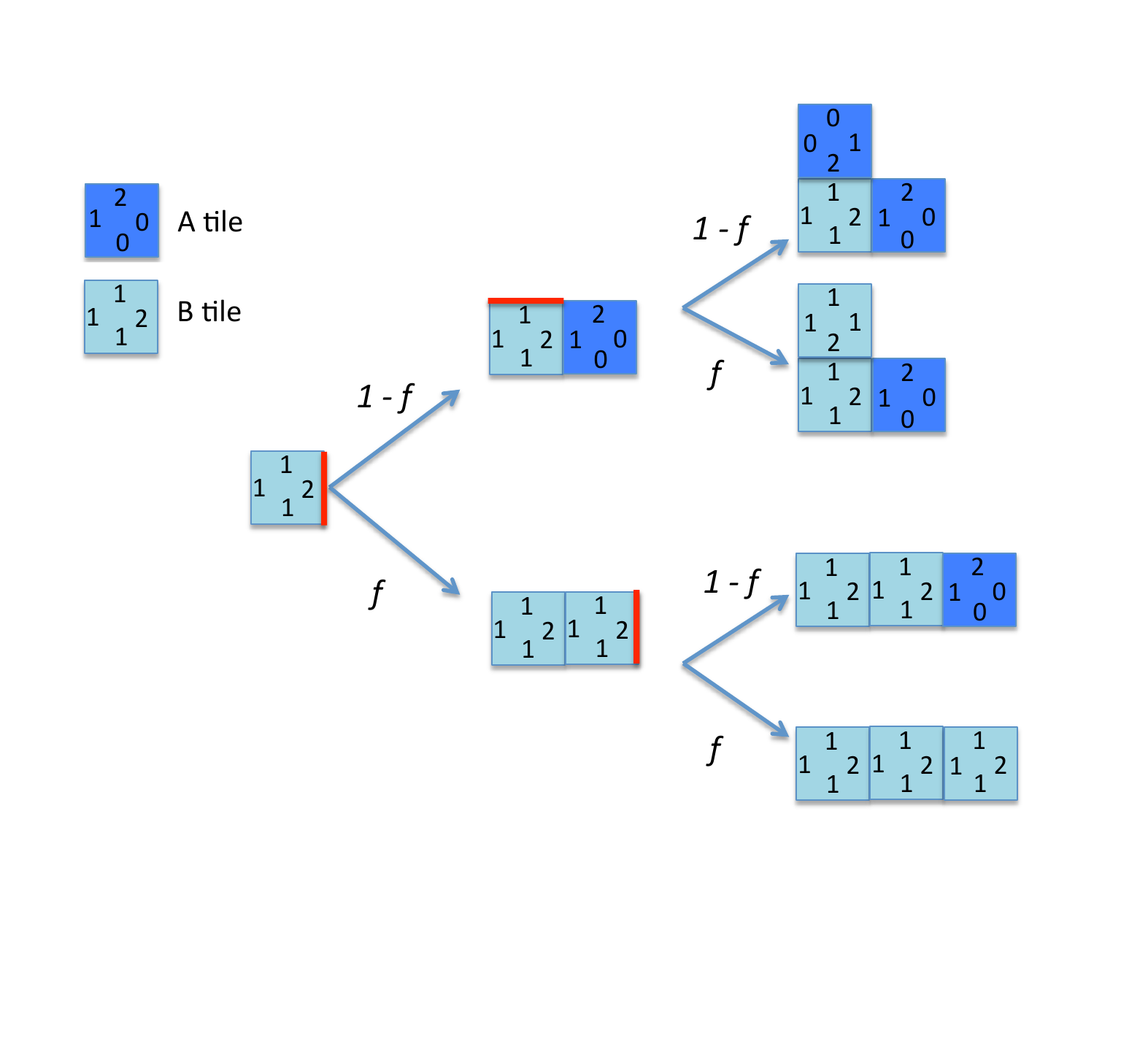}
\vspace{-2cm}
  \caption{Probability tree of possible evolutions of a structure grown with the two-tile set $\{1,2,0,0\}$-$\{1,2,1,1\}$ with asymmetric interactions. The perimeter site chosen for attachment in the next time step is marked in red.}\label{algorithm}
\end{figure}

In the following we describe a particular version of the lattice model of self-assembly introduced in \cite{Modularity}. The assembled structures are sometimes referred to as polyominoes \cite{tiling,math2}. Tiles are squares on a two-dimensional lattice, and the four sides of each tile are coloured. The colours represent possible (non-)interactions, and the set of coloured tiles and interaction rules is sometimes termed an `assembly kit' \cite{Modularity}. The colourings characterising the tiles are represented by integer numbers 0, 1 and 2. Colour 0 is neutral (meaning it does not interact at all), while colours 1 and 2 can interact with each other. The interaction is short-ranged and infinite in strength, causing tiles to bind permanently when interacting sides come in contact with each other. The two colours 1 and 2 can interact in two ways: (a) symmetrically, meaning that both colours bind to faces of the same colour, but not to the other colour, or (b) asymmetrically, meaning that colour 1 always binds to colour 2, but neither binds to itself. In the space of colours these two possibilities can be written in terms of binary interaction matrices, one symmetric ($I_S$) and the other asymmetric ($I_A$):
\begin{equation}
I_{S} =
  \left(
  \begin{array}{cc}
  1 & 0 \\
  0 & 1
  \end{array}
  \right),
  I_{A} =
  \left(
  \begin{array}{cc}
  0 & 1 \\
  1 & 0
  \end{array}
  \right)
\end{equation}
We denote the tile colourings in clockwise order for the four faces as, e.g. $\{1,2,0,0\}$. For two-tile sets we write e.g. $\{1,2,0,0\}-\{1,2,1,0\}$ and refer to the two tiles as A and B respectively. The assembly process starts with a seed tile, and then undergoes the following iterative steps: (i) In the case of two-tile sets, the tile type is chosen with probabilities $1-f$ (type A) and $f$ (type B) - note that $f$ remains fixed throughout a given assembly process. (ii) A rotational orientation is randomly chosen for the selected tile. (iii) Attachment of the tile is attempted with the selected orientation by randomly selecting a 'free' (i.e. interacting and unoccupied) side on the tiles that have already been assembled (i.e. in the first instance, the seed tile). (iv) If the sides in contact interact with each other, the structure grows. If not, the tile with the selected orientation is rejected. In both cases the assembly process continues iteratively with step (i). An example of possible assembly steps for the self-assembly tile set $\{1,2,0,0\}$-$\{1,2,1,1\}$ with asymmetric interactions is illustrated in Figure \ref{algorithm}. Growth ends when there are only neutral (i.e. colour 0) sides on the perimeter of the assembly. 

Observe that all free sides on the perimeter of the growing structure are equally likely to take part in the next assembly step, and that therefore empty lattice sites surrounded by more than one free face are more likely to be filled in a given assembly step. Selecting a free site with uniform probability simplifies the computational methods involved. We discuss a proposed experimental realisation of this assembly process in the discussion section of this paper. Our approach differs from models of other physical assembly processes, such as Diffusion Limited Aggregation \cite{DLA1}, because outer arms or edges in the growing structure are not more likely to be attached by new sides. Note that the self-assembly procedure defined above mirrors that used in \cite{Evo,Modularity,green} to study deterministic self-assembly. 

Self-assembly tile sets of the kind described above can exhibit a variety of behaviours, often dependent on the value of $f$. The structures produced can be either bound or unbound, and the assembly process can be deterministic or non-deterministic. These properties are defined as follows:
\begin{itemize}
 \item Unbound structures: those structures that will continue to grow indefinitely with a non-zero probability as $t\rightarrow \infty$, where $t$ is the total number of assembly time steps.
\item Bound structures: those structures for which assembly will terminate with unit probability (i.e. run out of free active sites for attachment) as $t \rightarrow \infty$.
\item Deterministic assembly: the assembly process always forms the same final structure (i.e. with the same tile type positions and orientations) as $t \rightarrow \infty$.
\item Non-deterministic assembly: the assembly process may lead to different final structures as $t \rightarrow \infty$.
\end{itemize}

In the following sections we first classify the growth of single tiles in these terms before investigating mixtures of two tiles. 

\section{Single-tile self-assembly}
\begin{figure}[]
\includegraphics[width=0.42\textwidth]{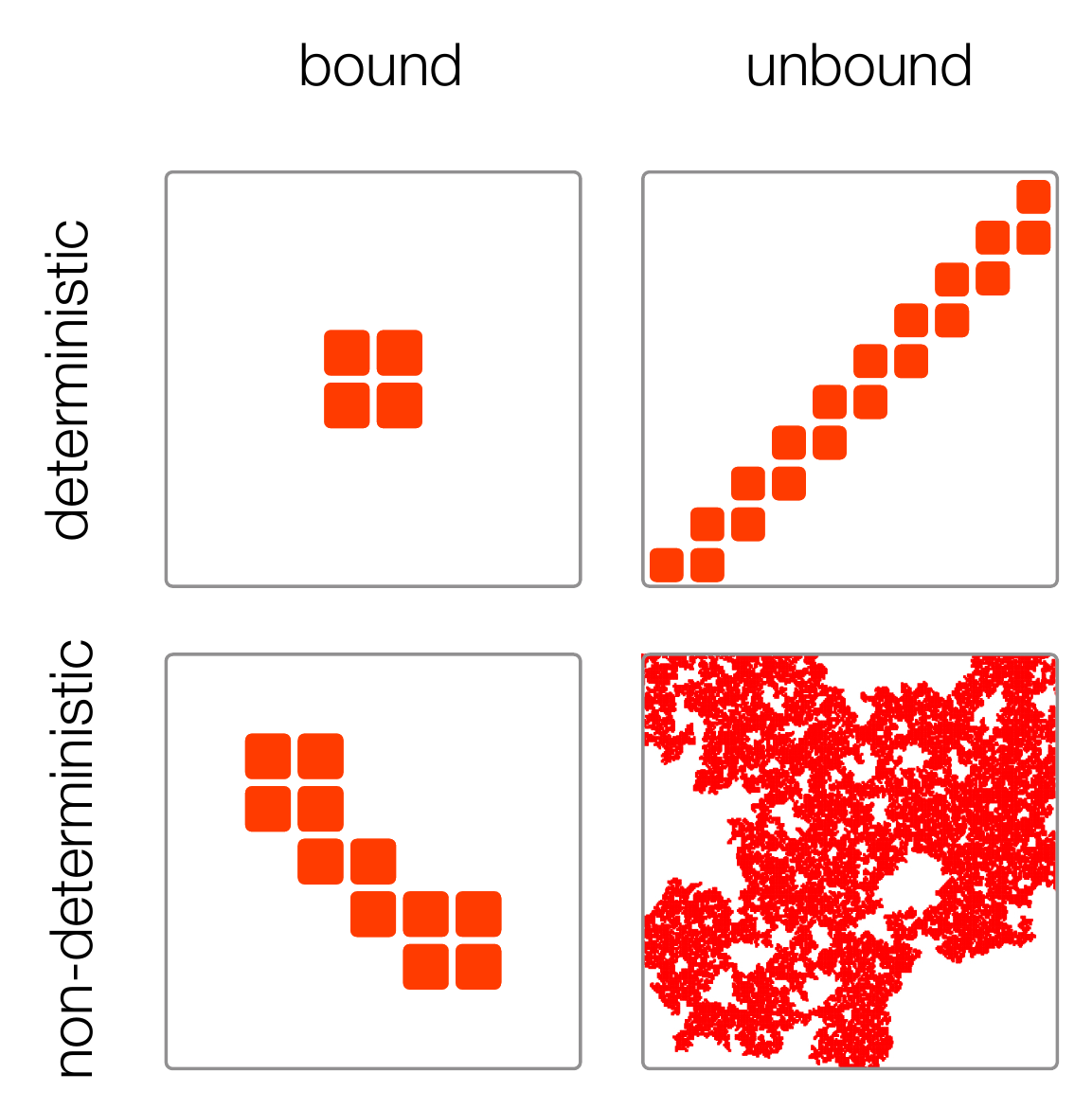}
\caption{The four different single tile behaviours that lead to growth: bound deterministic, unbound deterministic, bound non-deterministic, and unbound non-deterministic.}\label{budn} 
\end{figure}

Our investigation begins with exploring the growth mechanism with a single tile type. Seeding the assembly with an initial tile, the attachment of a tile of the same type, with random orientation, is attempted at each assembly time step. There are 13 topologically distinct interactive tiles that can be written down with the 0,1 and 2 colours. These correspond to the total number of necklaces of length four with up to three different colours \cite{Modularity,math}, taking into account additionally exchange symmetry of the colours 1 and 2, as swapping these labels will not change assembly. Table \ref{singletile} shows how the four single-tile growth behaviours outlined above are distributed among these single tiles, both for asymmetric and symmetric interactions. Figure \ref{budn} illustrates examples of each of these behaviours. Note that tiles with a single colour will not interact at all with each other under asymmetric interactions, which means that these tiles show no growth behaviour.
 
\begin{table}[]
  \centering
  \begin{tabular}{m{2.8cm} m{2cm} m{2cm}}
    \toprule
    \bf \center{Behaviour} & \multicolumn{2}{c}{\bf Interactions} \cr
    \bf & \bf \center{Asymmetric} & \bf \center{Symmetric}  \cr\cr
    \hline\cr
    \center{Bound  deterministic} & \center{$\{1,2,0,0\}$} & \center{$\{1,0,0,0\}$}  \cr\cr
    \hline\cr
    \center{Unbound  deterministic} & \center{ $\{1,2,1,2\}$ $\{1,0,2,0\}$} & \center{$\{1,0,2,0\}$ $\{1,0,1,0\}$ $\{1,2,0,0\}$ $\{1,2,1,2\}$ $\{1,1,1,1\}$}  \cr\cr
    \hline\cr
    \center{Bound non-deterministic} & \center{---} & \center{$\{1,1,0,0\}$}  \cr\cr
    \hline\cr
    \center{Unbound non-deterministic} & \center{$\{1,2,1,0\}$ $\{1,1,2,0\}$ $\{1,2,1,1\}$ $\{1,1,2,2\}$ $\{1,2,2,0\}$} & \center{$\{1,1,1,0\}$ $\{1,2,1,0\}$ $\{1,1,2,0\}$ $\{1,2,2,0\}$ $\{1,1,2,2\}$ $\{1,2,1,1\}$} \cr\cr
    \hline\cr
    \center{Non-interacting} & \center{$\{1,0,0,0\}$ $\{1,1,0,0\}$ $\{1,1,1,0\}$ $\{1,1,1,1\}$ $\{1,0,1,0\}$} & \center{---} \cr\cr 
  \bottomrule
  \end{tabular}
  \caption{The behaviours of all 13 topologically distinct single tiles, under symmetric and asymmetric interactions. Note that tiles with a single colour cannot bind to themselves under asymmetric interactions and thus cannot be classified as bound or unbound and deterministic or non-deterministic. Such tiles are therefore termed 'non-interacting'.}\label{singletile}
\end{table}
  
%\begin{table}[]

\section{Two-tile self-assembly}
We now consider self-assembly with two tile types. As we show below, the behaviours of two-tile sets can be classified in terms of the dependence of the final structure on the value of $f$, or in other words, the relative concentrations of the two tile types. In order to distinguish different types of $f$-dependence we plot the density of the assembled structure in a pre-defined region, as a function of $f$. As most of these two-tile assembly processes result in roughly isotropic growth in two dimensions, the pre-defined region is a circle of radius ${1 \over 2} \sqrt{{N \over \pi}}$ centred around the seed tile. The factor of ${1 \over 2}$ means that the outer perimeter of the structure is highly likely to lie outside the region of the density measurement. Furthermore, as we show below, the fact that the growth behaviour of some self-assembly tile sets is highly anisotropic can also be captured using this density measurement.  

\subsection{Critical behaviours}

We define {\em critical behaviour} of a two-tile self-assembly tile set as the existence of a critical value of $f$ below (or above) which structures are bound, and above (or below) which structures are unbound. This behaviour results in a density function with an inflection point as an unbound structure leads to the saturation of the measurement region. 
\begin{figure}[]
 \centering
 \includegraphics[width=0.6\textwidth]{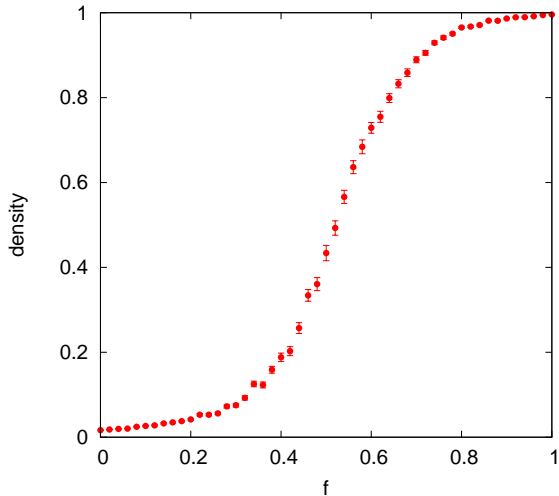}
 \caption{The density function for the tile set $\{1,2,0,0\}$-$\{1,2,1,0\}$. The inflection point signals the change in behaviour from bound to unbound.}\label{sca}
\end{figure}

\begin{figure}[]
  \includegraphics[width=0.5\textwidth]{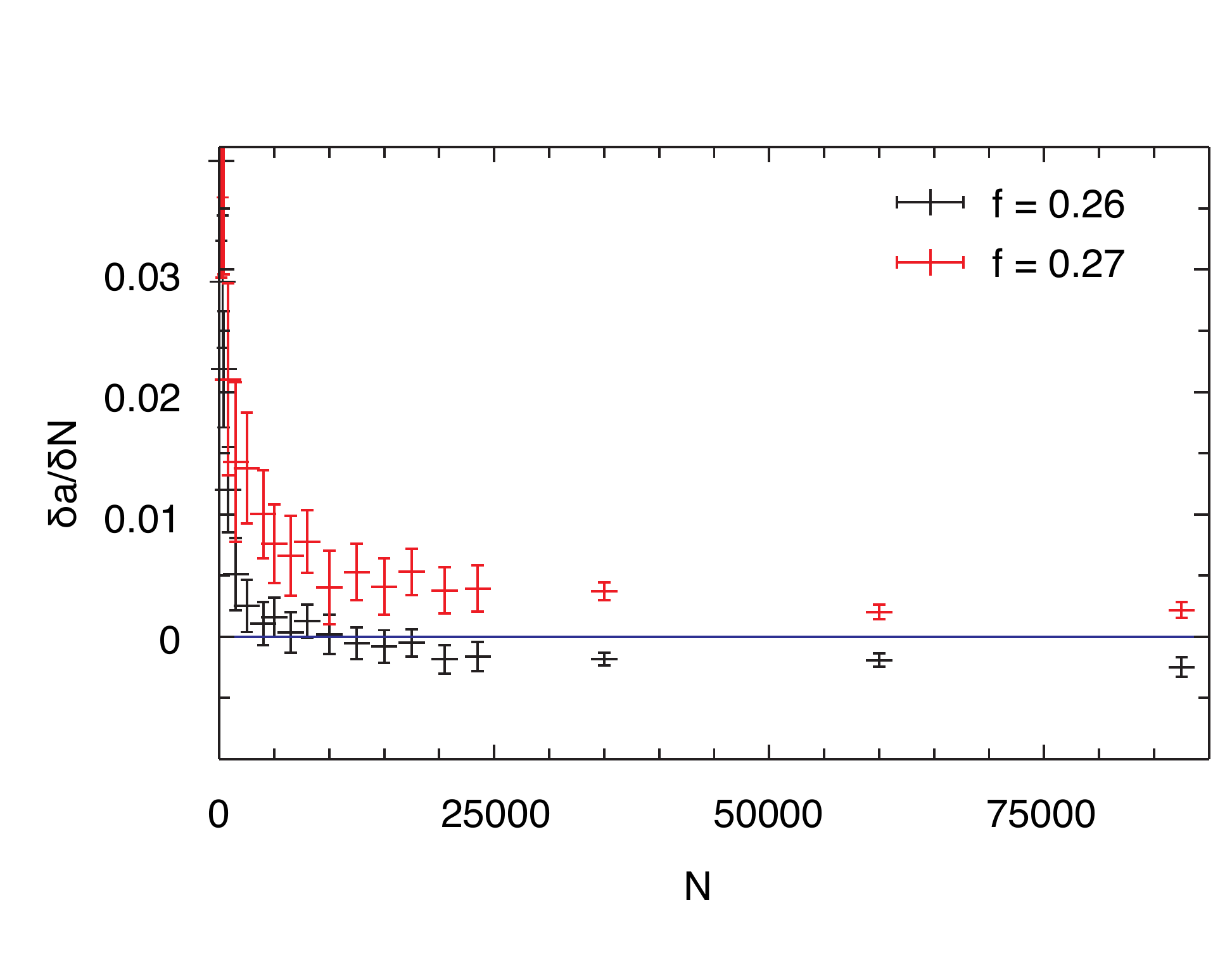}
  \caption{The rate of change in the number of free sides ${\delta a }\over{\delta N}$ for the tile set $\{1,0,0,0\}$-$\{1,1,1,1\}$, and for a value of $f$ around the critical value of $f_c$. At large $N$ the growth rate is negative for $f = 0.26$ and positive for $f = 0.27$, which means that $0.26 < f_c < 0.27$.}\label{fdelta2}
\end{figure}

Figure \ref{sca} shows the density function for set $\{1,2,0,0\}$- $\{1,2,1,0\}$ with asymmetric interactions. According to the definitions presented above, a structure is unbound if it has a non-zero probability of continuing to grow forever. We define the critical value $f_c$ in terms of the average rate of change in the number of free sides $a$ as the structure grows, in other words ${\delta a} \over {\delta N}$, over a large number of assembly runs. For $f = f_c$, the value of ${\delta a} \over {\delta N}$ for large $N$ changes sign - an example is shown in Figure \ref{fdelta2} for the self-assembly tile set $\{1,0,0,0\}$-$\{1,1,1,1\}$. In this case for values of $f$ above $f_c$ the number of free sides continues to increase, leading to unbound growth. In the remainder of this paper we define $f_c$ as the value for which ${{\delta a} \over {\delta N }} = 0$ at $N=50000$. We choose this value because we observe, in extensive simulations, that the changes in ${\delta a} \over {\delta N}$ with increasing N typically approach zero around $N = 10000$, representing a steady state plateau of ${\delta a} \over {\delta N}$. This state is very well established by $N = 50000$ in all our simulations, making it possible to determine $f_c$ according to the above definition.

\begin{figure}[h]
  \centering
  \includegraphics[width=0.5\textwidth]{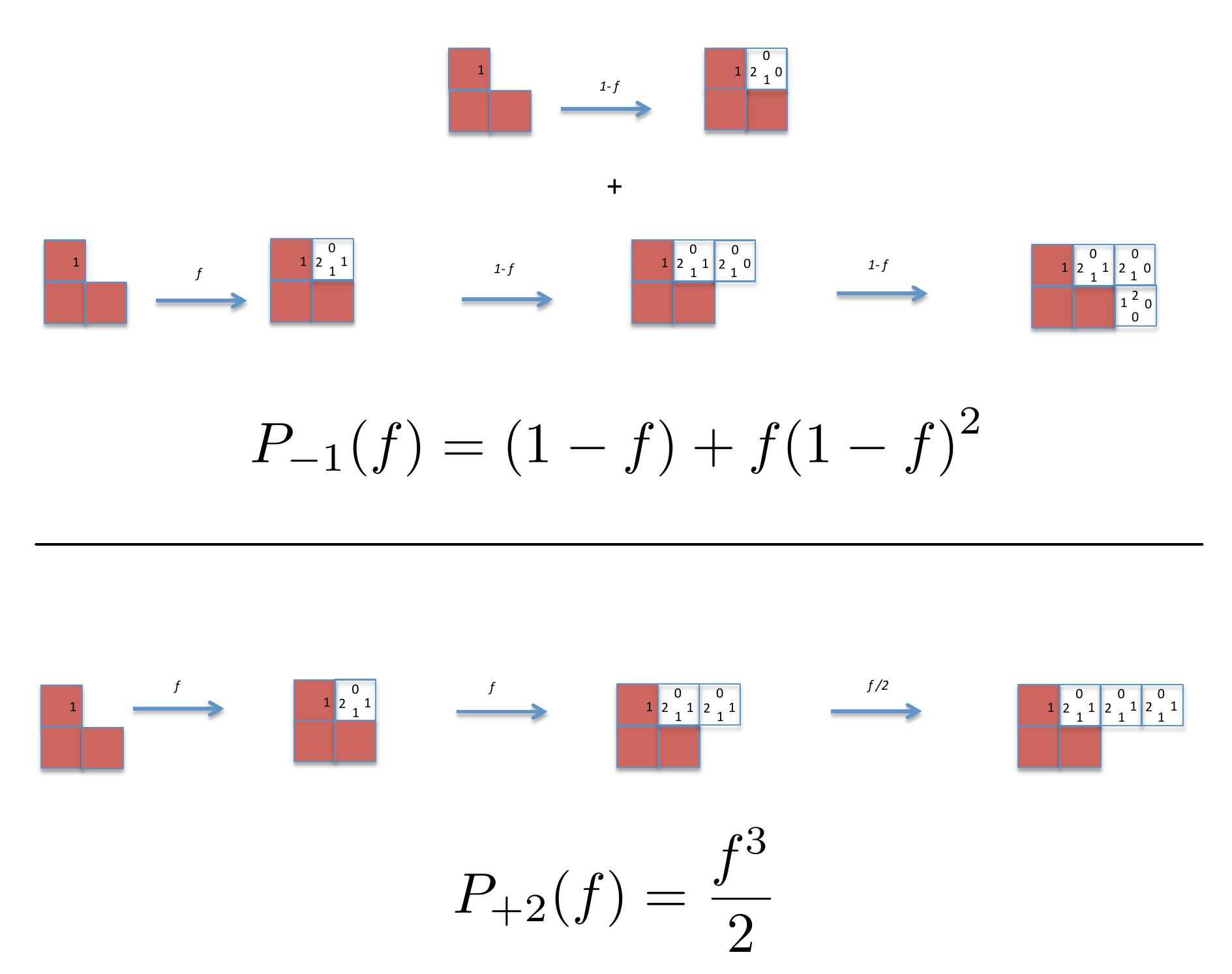}
  \caption{Graphical representation of two of the terms arising from equation \ref{no} for tile-set $\{1,1,2,0\}$-$\{1,2,0,0\}$ with asymmetric interactions. The terms shown are for $\delta a = -1$ (top) and $\delta a = +2$ (bottom). This means that the final structures in the top two rows have zero free faces, and therefore one less ($\delta a = -1$) than the initial corner configuration, and that the final structure in the bottom row has three free faces, two more ($\delta a = +2$) than the initial configuration.}\label{algorithm2}
\end{figure}

In total there are 18 critical tile sets with symmetric interactions, and 9 critical tile sets with asymmetric interactions (not counting chiral duplicates). Predicting the critical value $f_c$ from the tile colourings alone is a difficult task. We present here an approach that provides a reasonable estimate of $f_c$ for a subset of all possible two-tile sets displaying critical behaviour, namely (i) all such sets with a single colour and symmetric interactions, as well as (ii) all such sets with two colours and asymmetric interactions in which one colour (the colour 2, without loss of generality) does not appear more than once on either tile. The consequence of these restrictions is that there is at most one free side of colour 2 on the structure and typically many free sides of colour 1. The calculation of our estimate is based on the relative addition or subtraction of free sides over three successive time steps, starting with a common intermediate local assembly environment in all of the above assembly sets. This common environment is a corner site with a free side of colour 1. The reason for choosing this particular scenario is that flat regions of the perimeter of a structure almost inevitably contain free sides which will grow such corners at a later stage of the assembly. The corners however can be precursors of growth termination, which is why they represent a useful starting point. We furthermore consider corners of a fixed chirality, in which the interacting side is preceded (clockwise) by a non-interacting side. Since assembly sets can have a chirality, and since we are interested only in the overall behaviour rather than the chirality of the structure, we are free to choose the chirality of our sets such that the corner defined above is more likely to lead to growth termination. This is the case if we ensure that the set chirality is chosen such that the set contains tile $\{1,2,0,0\}$ rather than $\{2,1,0,0\}$ (where appropriate). See Figure \ref{algorithm2} for an illustration. 

We now consider the assembly of up to three tiles, starting with the local corner environment described above. The focus on a limited number of tiles allows us to write down an analytical expression for the mean number $\overline{\delta a}(f)$ of free faces gained or lost:

\begin{equation}\label{no}
\overline{\delta a}(f) = \sum P_{\delta a}(f)\delta a = 0
\end{equation}

where $P_{\delta a}(f)$ is the probability of a change $\delta a$ in the number of free faces, for a given $f$-value. The solution of $\overline{\delta a}(f) = 0$ is $f = f_c$.

Table \ref{table2} gives the analytical expressions for the mean change in the number of free faces as a function of $f$ for the eleven tile sets that fall into categories (i) and (ii) outlined above. The solutions of these equations set to zero provides the values of $f_c$.

\begin{figure}[h]
\centering
\includegraphics[width=0.6\textwidth]{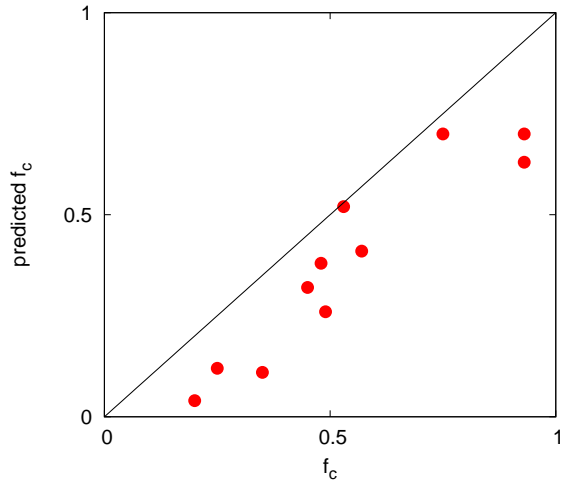}
\caption{Predicted versus actual values of $f_c$ for the critical self-assembly tile sets listed in Table \ref{table2}. Note that the predicted values all lie lower than the actual ones. The likely reason for this is that larger-scale steric effects are not taken into account in this model. The correlation coefficient is 0.935.}\label{scale}
\end{figure}

The predictions show a strong correlation with the actual values (Pearson correlation 0.935), but all of the predicted values of $f_c$ are lower than the real $f_c$, as shown in Figure \ref{scale}. This is likely to be the result of growth termination due to larger-scale effects, such as the growth of a larger part of the structure into itself. Note that predictions are worse for highly chiral sets like $\{1,1,2,0\}$-$\{1,2,0,0\}$, where growth of active sites is highly correlated.

\begin{table*}[]
 \begin{tabular}{c|c|c|c}
  \multicolumn{4}{c}{\bf Symmetric interactions}\\[3pt]   
  \hline
&&&\\[-6pt]
  tile set & $f_c$& predicted $f_c$  & mean change in number of faces, $\overline{\delta a}(f)$    \\[3pt]
  \hline
&&&\\[-6pt]

  $\{1,0,0,0\}$-$\{1,1,1,1\}$ &  $ 0.25 $ & 0.12 & $236 f^3+72 f^2-3 f-1$ \\[3pt]
  \hline                       
&&&\\[-6pt]

  $\{1,1,0,0\}$-$\{1,1,1,1\}$ &  $ 0.2 $& 0.04 &  $\frac{1}{3} \left(321 f^3+463 f^2+134 f-6\right) $  \\[3pt]
  \hline 
&&&\\[-6pt]

  $\{1,0,0,0\}$-$\{1,1,1,0\}$ &  $ 0.49 $& 0.26 & $\frac{152 f^3}{3}+10 f^2-2 f-1 $ \\[3pt]
 \hline 
&&&\\[-6pt]
  $\{1,1,0,0\}$-$\{1,1,1,0\}$ &  $ 0.35 $& 0.11 & $\frac{1}{6} \left(77 f^3+194 f^2+87 f-12\right) $  \\[3pt]
 \hline
\end{tabular}
\vspace{1cm}

 \begin{tabular}{c|c|c|c}
 \multicolumn{4}{c}{\bf Asymmetric interactions}\\[3pt]
  \hline
&&&\\[-6pt]
  tile set  & $f_c$& predicted $f_c$ & mean change in number of faces, $\overline{\delta a}(f)$    \\[3pt]
\hline
&&&\\[-6pt]
 {\bf $\bf \{1,2,0,0\}$-$\bf\{1,2,1,0\}$} &  $ 0.53 $ & 0.52 & $-\frac{f^3}{2}+2 f^2+f-1 $ \\[3pt]
  \hline 
&&&\\[-6pt]
  {\bf $\bf \{1,2,0,0\}$-$ \bf\{1,2,1,1\}$} &  $ 0.45 $& 0.32 & $1/8 (-8 + 14 f + 33 f^2 - f^3) $\\[3pt]
  \hline 
&&&\\[-6pt]
  {\bf $\bf\{1,2,0,0\}$-$\bf\{2,1,1,0\}$ }&   $ 0.48 $& 0.38 & $ - \frac{5 f^3}{12} + \frac{5 f^2}{2} + \frac{7f}{4}-1 $\\[3pt]
  \hline
&&&\\[-6pt]
  {\bf$\bf\{1,2,0,0\}$-$\bf\{1,1,2,0\}$ }&  $ 0.93 $& 0.63 & $\frac{5 f^2}{2}-1 $\\[3pt]
  \hline
&&&\\[-6pt]
      {\bf $\bf\{2,0,0,0\}$-$\bf\{1,1,2,0\}$} &  $ 0.93 $& 0.7 &$\frac{3 f^3}{2}+f^2-1 $\\[3pt]
 \hline    
&&&\\[-6pt]
 $\{2,0,0,0\}$-$\{1,2,1,0\}$ &  $ 0.75 $& 0.7 & $-1 + f^2 + (3 f^3)/2 $\\[3pt]
  \hline                       
&&&\\[-6pt]
 $\{2,0,0,0\}$-$\{1,2,1,1\}$ & $ 0.57 $ &0.41 & $-\frac{f^3}{4}+6 f^2-1 $\\[3pt]

  \specialrule{.1em}{.05em}{.05em} 
\end{tabular}
 \caption{Predictions of $f_c$, the critical value of $f$, for the critical self-assembly tile sets with a single interactive colour under symmetric and asymmetric interactions. The predicted value of $f_c$ is the real root on the unit interval of the equation $\overline{\delta a}(f) = 0$. This condition represents the value of $f$ at which the number of active sites on the perimeter of the assembled structure no longer grows in the long term. Note that tile sets possessing chiral duplicates have been marked in bold, and their chiral duplicate has been omitted.}\label{table2}
 \end{table*}
 
The self-assembly tile sets that exhibit critical behaviour also display self-similar structures at criticality. In particular we observe two fractal classes, one with fractal dimension $1.32 \pm 0.05$ and one with fractal dimension $1.2 \pm 0.03$. The value of the fractal dimension depends on the chirality of the `unbound' tile in the tile set, i.e. the tile that on its own would give rise to an unbound assembly according to the classification of single-tile behaviours in Table \ref{singletile}. In the context of this paper, the Hausdorff box-counting dimension is taken as the fractal dimension of our structures \cite{1918}. All critical tile sets, along with all other self-assembly tile sets, are presented in the supplementary tables S1-S4 with their critical values $f_c$ and fractal dimensions. Aside from the critical transitions discussed in this section, numerous other behaviours can be observed for two-tile sets, which are described in the following sections.

\subsection{Dimensional transitions}\label{sal}
\begin{figure}[]
  \centering
   \includegraphics[width=0.6\textwidth]{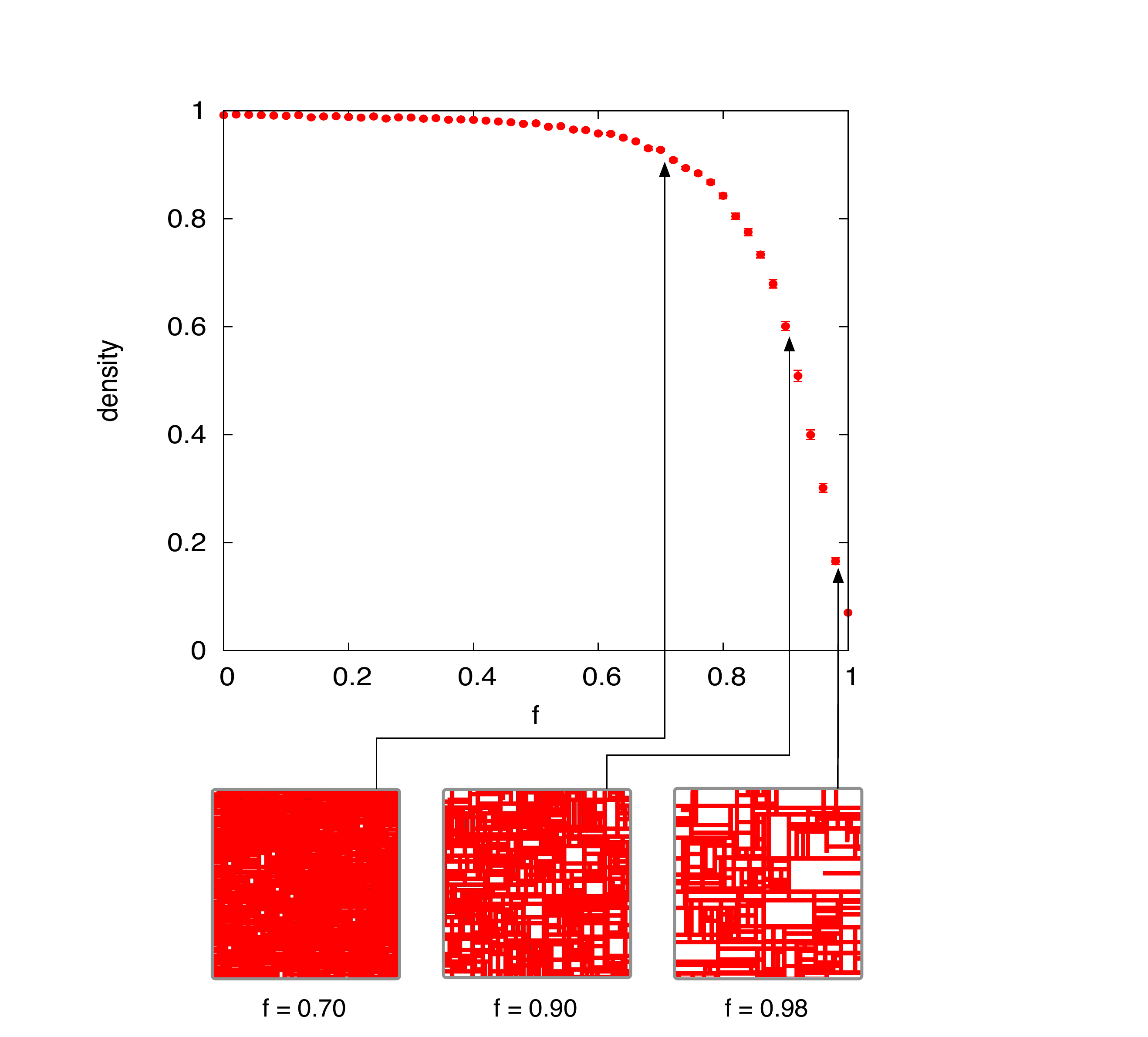}
  \caption{Density transition from dense to dilute patterning with tile set $\{2,1,2,1\}$ - $\{2,0,1,0\}$ with asymmetric interactions. }\label{patt}
\end{figure}

Figure \ref{patt} shows evidence for a non-critical transition in structural density, in form of a monotonic density function without an inflection point. The tile $\{1,0,2,0\}$ under asymmetric interactions (which grows in a straight line) dilutes the isotropic growth of tile $\{1,2,1,2\}$. This results in isotropic growth in two dimensions of varying density, apart from when $f=1$, when the growth is strictly one-dimensional. We term this a `dimensional transition'. 

Using the density function we classify 30 self-assembly tile sets as dimensional transitions: 24 with symmetric interactions and 6 with asymmetric interactions.

\subsection{Random-walk assemblies}
\begin{figure}[h]
  \centering
    \includegraphics[width=0.6\textwidth]{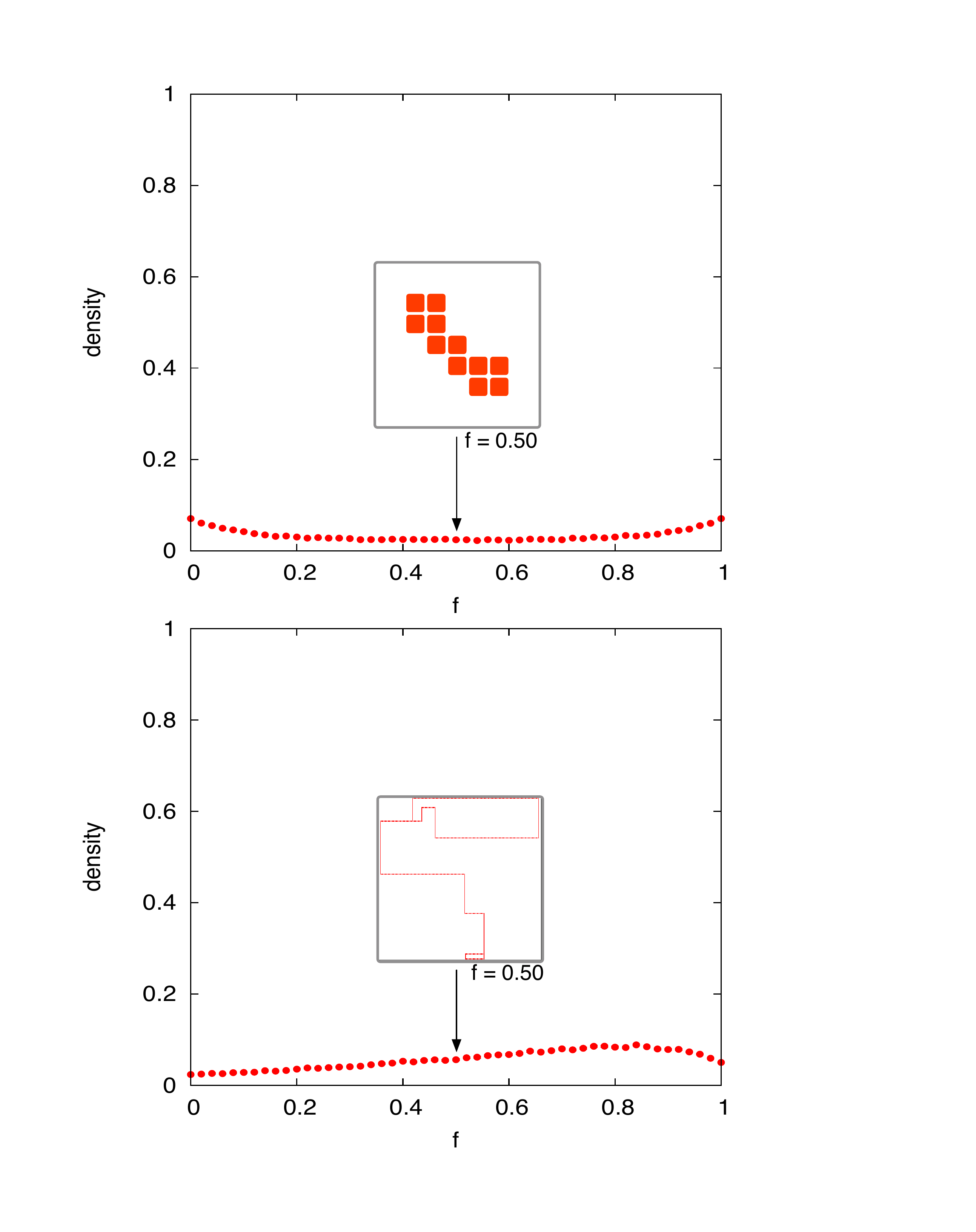}\\
  \caption{Top: Examples of density functions for tile sets $\{2,1,0,0\}$-$\{1,2,0,0\}$ (A) and $\{1,1,0,0\}$-$\{1,0,1,0\}$ (B), both under symmetric interactions, that result in random walk-assemblies. Bottom: Structures grown at $f=0.5$.}\label{RWW}
\end{figure}

For some self-assembly tile sets the total number of free sides $a$ on the structure can never increase beyond the number of interactions on a single tile. The structures assembled by such tile sets resemble a random walk motion on a plane with a turning probability that depends on $f$. For instance, for the tile set $\{1,1,0,0\}$-$\{1,0,1,0\}$ with symmetric interactions, structures stop growing when the the two `heads' of the walker bite their `tail'. In the case of random-walk assemblies the density function typically shows a low, non-monotonic density for varying $f$, as shown in Figure \ref{RWW}.

\subsection{Bottleneck growth}
\begin{figure}[]
  \centering
    \includegraphics[width=0.6\textwidth]{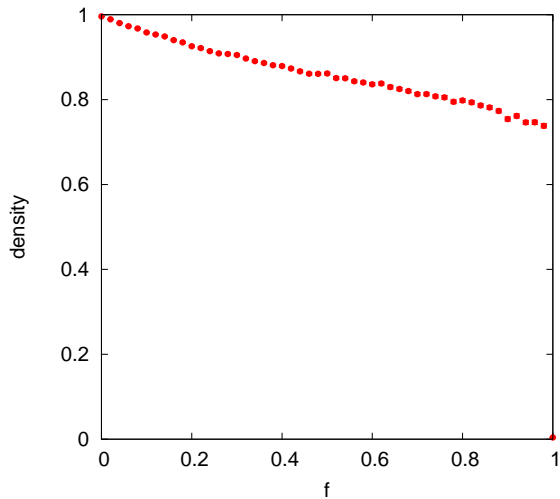}
  \caption{The density function for tile set $\{1,2,1,0\}$-$\{0,2,0,2\}$ under asymmetric interactions exhibits a linear transition in density as a result of $f$-dependent bottleneck growth.}\label{patt3}
\end{figure}

\begin{figure}[]
  \centering
 \includegraphics[width=0.6\textwidth]{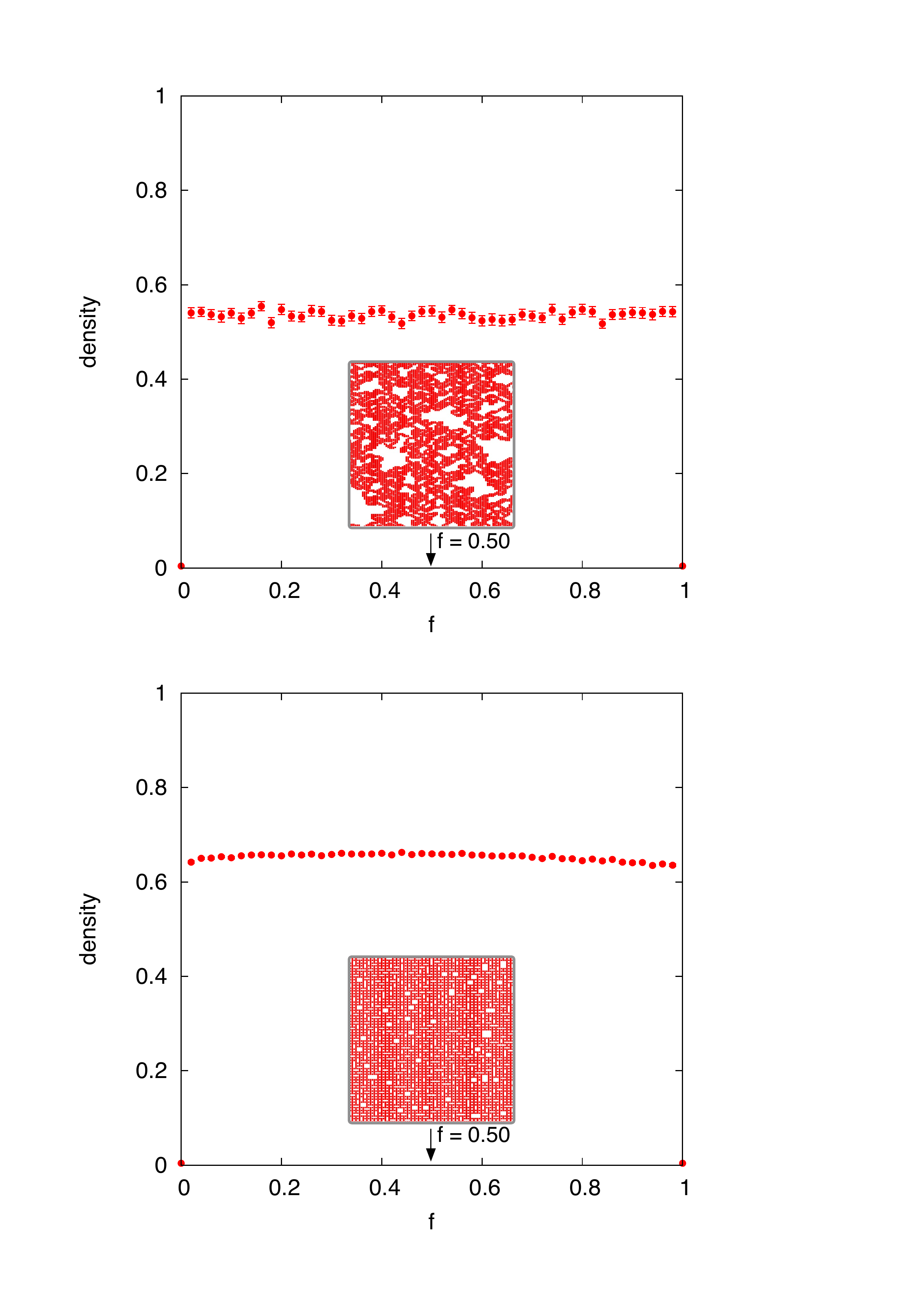}
  \caption{TOP: The density functions for the tile sets $\{2,2,0,0\}$-$\{1,1,1,0\}$ and $\{2,0,2,0\}$-$\{1,1,1,0\}$, both under asymmetric interactions, exhibit $f$-independent behaviour. BOTTOM: Structures grown at $f=0.5$}\label{11}
\end{figure}

Another type of non-critical density transition can arise when one tile cannot interact with itself, and is therefore dependent on the other tile. This gives rise to a linear change in density with $f$, as shown in Figure \ref{patt3} for the tile set $\{1,2,1,0\}-\{0,2,0,2\}$ with asymmetric interactions. We term this behaviour $f$-dependent bottleneck growth. This contrasts with $f$-independent bottleneck growth, which means that the tiles interact with each other but neither tile interacts with itself. In this case the density is constant. Two rather different examples can be seen in Figure \ref{11} for tile sets $\{2,2,0,0\}$-$\{1,1,1,0\}$ and tile set $\{2,0,2,0\}$-$\{1,1,1,0\}$, both under asymmetric interactions. 

\subsection{High-density non-monotonic growth}\label{D1}
\begin{figure}[h]
  \centering
    \includegraphics[width=0.6\textwidth]{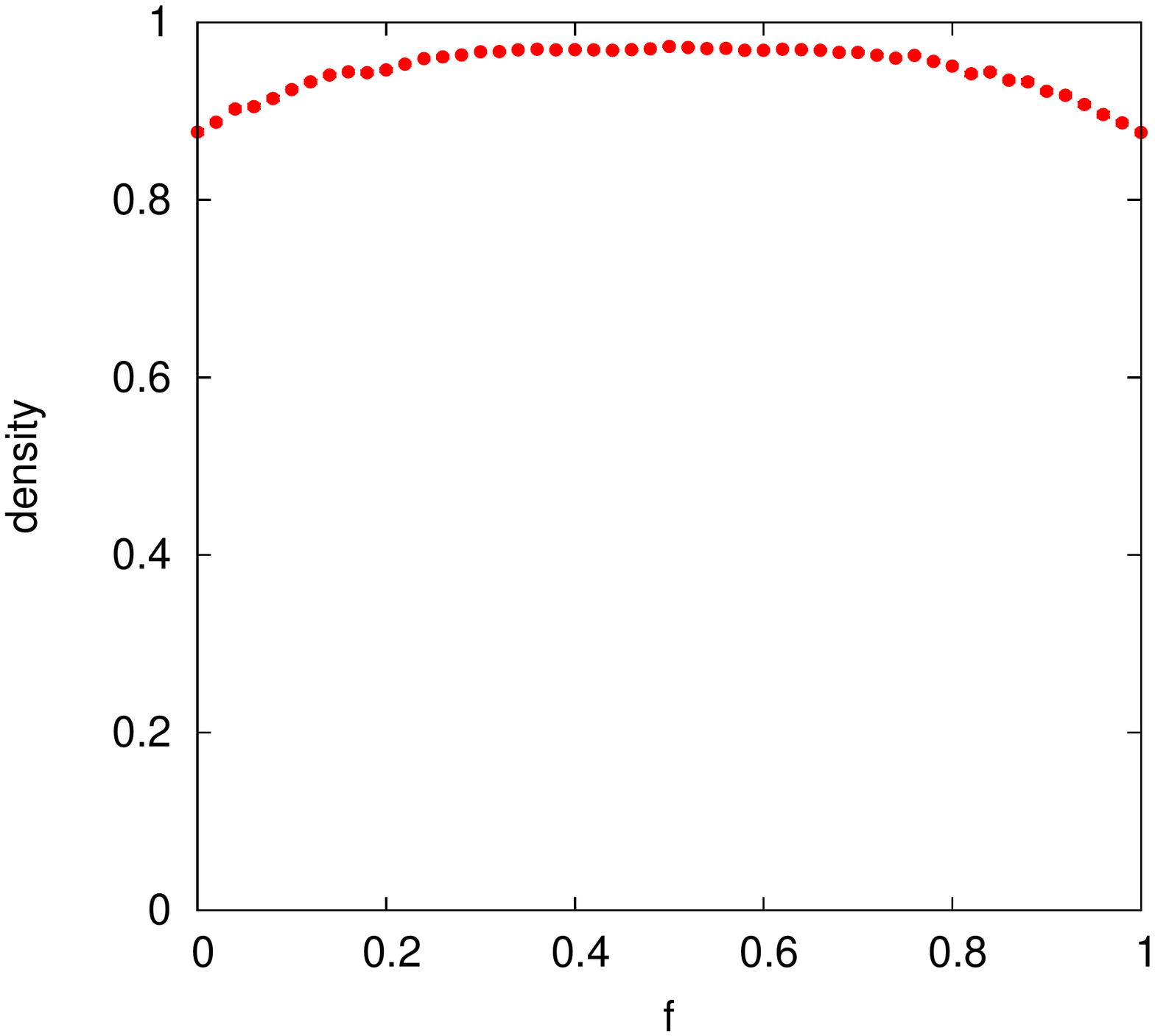}
  \caption{Tile set $\{2,1,2,0\}$-$\{1,2,1,0\}$ with symmetric interactions shows a non-monotonic and overall high density with are larger hole presence for high and low $f$ values.}\label{112}
\end{figure}

In Figure \ref{112} tile set $\{2,1,2,0\}$-$\{1,2,1,0\}$ with symmetric interactions is shown. The non-monotonic density function reveals that there is a more abundant presence of holes in the structure for high and low $f$. However, structures stay unbound and non deterministic for all $f$.

\subsection{Seed-dependent assemblies}\label{D3}
\begin{figure}[]
  \centering
  \includegraphics[width=0.6\textwidth]{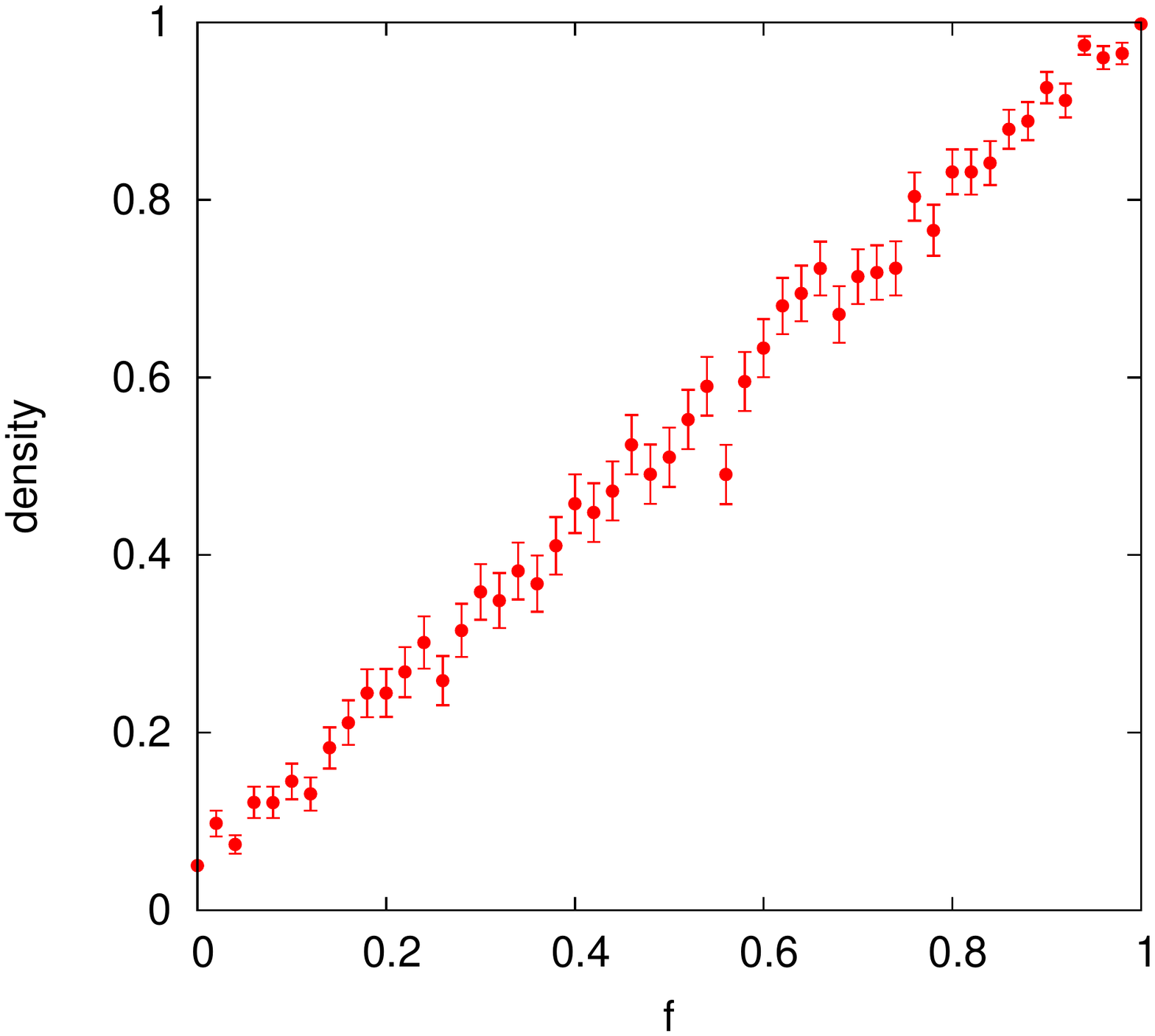}\\
 \caption{An example of a seed-dependent transition, for tile set $\{2,0,2,0\}-\{1,1,1,1\}$ under symmetric interactions.}\label{patt5}   
\end{figure}

Some assemblies are strongly dependent on seed choice. For example the tile set $\{2,0,2,0\}$-$\{1,1,1,1\}$ with symmetric interactions can build two very different structures, depending entirely on which of the tiles is picked at the seed stage. The parameter $f$ governs the probability of that initial choice, but has no influence after that. Since the density functions are constructed using a large number of repeated assembly runs, the density is simply $(1-f) \rho_A + f \rho_B$. For the example above this is approximately $f$ (see Figure \ref{patt5}), since the density of $\{2,0,2,0\}$ is close to zero and $\{1,1,1,1\}$ fills the plane. 

In a second class of seed-dependent behaviours, which we term `extended seed-dependence', the consecutive choice of the first two tiles determines the final behaviour of the self-assembly tile set. For example, the tile set $\{1,2,1,2\}$-$\{1,0,0,0\}$ with symmetric interaction can result in bound growth only if two $\{1,0,0,0\}$ tiles are chosen in the first two assembly steps. Otherwise the system will always be unbound.

\subsection{Tile set criteria}

Below follows a summary of the tile set criteria that define the different categories described in the previous sections. Note that critical behaviours and dimensional transitions are detected using the density functions and $da/dN$. 

\begin{itemize}
\item seed-dependence: the tiles must interact with themselves but not with each other.
\item extended seed-dependence: this occurs under symmetric interactions when one tile only has one single interaction, and the other colour appears at least twice, and on opposite faces, on the other tile. 
\item $f$-dependent bottleneck: one tile must interact with itself and with the other, but the other one does not interact with itself.
\item $f$-independent bottleneck: the tiles interact with each other but not with themselves.
\item HDNM: The tiles display more than three active faces each, and interact with themselves and with each other.
\item random walker: both tiles display at most two interactive faces and interact with each other. 
\end{itemize}

For a full table of the properties of all 106 tile sets see Supplementary Tables S1-S4. 

\section{Discussion}\label{discussion}
Self-assembly tile models, such as the one presented here, are highly abstract, but nevertheless provide a useful model for real assembly phenomena, particularly in a biological context. An example are protein aggregation and polymerization, which are the hallmarks of a number of diseases, including sickle-cell anaemia and a number of diseases related to amyloid fibrils. For sickle-cell anaemia a self-assembly tile model has already been developed \cite{green}, illustrating how a single point mutation can turn a tile set of two tile types from a bound deterministic structure into an unbound deterministic structure. Specifically, one can model the $\alpha$ and $\beta$ chains of wild-type haemoglobin Hb A as a $\{1,3,0,0\}-\{2,5,0,4\}$ tile set, and the sickle-cell mutant Hb S as the tile set $\{1,3,0,0\}-\{2,5,{\bf 6},4\}$ (mutation shown in bold) \cite{green}. Pairs of colours interact asymmetrically, meaning that 1 binds to 2, 3 to 4, and 5 to 6, so that each colour only binds to one other specific colour. The mutation in the Hb S tile set corresponds to the mutation of the sixth residue in the $\beta$ chain from glutamic acid (E) to valine (V), which causes a hydrophobic patch that binds to a pocket between residues 85-88 on the same $\beta$ chain, and causes polymerisation of Hb S \cite{Bunn:1997de}. The pocket of residues is represented by the colour 5 on the tile set. As this colour can only bind to the colour 6 that emerges in the mutant, it does not bind to anything in the wild-type tile set \cite{green}. This polyomino model of sickle-cell anaemia requires a larger number of colours than the tile sets investigated in this paper, which are limited to two non-neutral colours. Nevertheless it illustrates that highly simplified self-assembly tile models are capable of reproducing important characteristics of real protein polymerization and aggregation phenomena, and that further exploration of the space of assembly possibilities, for example by investigating tile sets with more colours or more tiles, would certainly be worthwhile.

Aside from protein aggregation and polymerisation our model also has potential implications for nanoscale engineering in the form of self-assembling DNA \cite{Seeman1983,winfree1998design,winalgo,rothemund2006folding,fujibayashi2007toward,Wei2012d} and RNA \cite{chworos2004building} tiles. This field has developed enormously over the past two decades, and many of the challenges that separated the idealised assembly from the reality of the experimental outcomes, such as the lack of tile rigidity \cite{winfree1998design,chworos2004building}, and a variety of binding and nucleation errors \cite{winalgo} have now largely been overcome \cite{rothemund2006folding,fujibayashi2007toward,Wei2012d}. One of the specific differences between our model and DNA tiles is that most DNA self-assembly experiments start with multiple seed tiles, as opposed to a single seed tile. The single-seed model can however also be implemented, if we are able to carefully control the availability of the tiles. The implications of the difference between single- and multi-seed assembly are the subject of an ongoing experimental and computational investigation by us (to be published separately). Another difference between our model and real DNA tiles is the rigidity of DNA assembly blocks. While significant improvements have been made on this front, real scaffold-free DNA self-assembly tiles are still much less rigid than the square tiles of our model. However, first results in the ongoing experiments mentioned above seem to indicate that the critical behaviours we observe in our lattice model are largely unaffected by these differences in tile shape.

As discussed above the predictions of $f_c$ are lower bounds because larger-scale steric effects are not taken into account by our analytical model, which only takes into account three growth steps. The combinatorics of the growth possibilities prohibits an expansion of this approach much beyond three steps. For example, there are more than a thousand possible combinations of positions, orientations, and tile type choices if we consider up to three growth steps for tile set $\{1,1,0,0\}-\{1,1,1,1\}$. Despite its simplicity, this analytical procedure yields estimates that correlate well across the range of observed $f_c$ values (Pearson correlation: 0.935).

The observation of only two different fractal dimensions might suggest that a simple analytical derivation of these dimensions is possible. However as the large-scale growth effects are difficult to predict from local growth possibilities, such an analytical model remains out of reach for the time being.

On the surface the critical behaviour of some of the self-assembly tile sets resembles percolation phenomena. For example, one of the self-assembly tile sets we consider, namely $\{1,1,1,1\}$-$\{1,0,0,0\}$ with symmetric interactions, resembles the Eden model with blocking, which in turn can produce similar final structures to bond percolation \cite{surface,fractals111}. However, the growth process in a percolation model relies on the spatial independence of occupancy probabilities, whereas the self-assembly model has strong local correlations. 

In percolation theory we can compute the $\gamma$ exponent that describes the scaling of the average cluster size close to criticality, $\langle s \rangle \propto |p-p_c|^{\gamma}$, where $\sum s^2 n_s(p)= \langle s \rangle$. To do so, one needs to measure the size distribution of the percolation clusters $n_s(p)$ for lattices of varying size. It is impossible to realise the self-assembly phenomenon described here through an equivalent lattice model because our self-assembly clusters start from a single seed. Hence, a $\gamma$ exponent cannot be defined and calculated through finite size scaling arguments analogously to percolation theory \cite{cardy}. In other words, there is no way to define $n_s(p)$ with our self-assembly algorithm. 

Modifying our seeding conditions, other exponents could be defined such as ${\xi}_{\bot}$ and ${\xi}_{\parallel}$, but such extensions, based on models such as \cite{Frey1}, are beyond the scope of this paper.

A three-dimensional implementation of this self-assembly model would exhibit a different set of $f$-dependent self-assembly behaviours, particularly regarding random-walk assemblies, which are bounded in 2D, but not in 3D, as $N$ goes to infinity. Another difference is that two-dimensional holes can be filled in our model, whereas it would be unrealistic to allow this in three dimensions. However, we would still expect to see critical behaviours in 3D, such as a critical transition for a 3D cube-set with faces $\{1,0,0,0,0,0\}-\{1,1,1,1,1,1\}$. Like in the equivalent 2D tile set $\{1,0,0,0\}-\{1,1,1,1\}$, $f$=0 would represent a bound state, even though the steric effects would differ in 3D.

\section{Conclusion}
Despite its simplicity, the self-assembly of two tile types on a square lattice with up to two interactions yields a wide variety of complex assembly behaviours. Of particular interest are critical transitions, that typically show a sharp dependence on the relative concentrations of the two tiles. These are of interest in the context of disease-related protein aggregation and polymerization phenomena in biology, such as the formation of amyloid fibrils and sickle-cell anaemia (see Discussion section). While such models are likely to require more than two colours, and possibly more than two tile types, this study can serve as a starting point for modeling such phenomena using self-assembly tiles.   

There are many avenues for generalising this work, to three dimensions, and to larger numbers of tile types and colours. The example of a self-assembly tile model for sickle-cell anaemia (see above), which requires six non-neutral colours, demonstrates that such generalisations would be worthwhile. A comprehensive framework for predicting whether a tile set is unbound or bound, and non-deterministic or deterministic would also be very desirable. Progress on these generalisations is likely to be made by studying the possible topologies of graphs of interactions between tile types.

The analytical prediction of $f_c$ for the critical self-assembly tile sets is difficult to achieve from first principles (i.e. from the tile colourings), but we present a first approach that provides reasonable estimates for a subset of these tile sets. 

A physical realisation of the self-assembly processes described in this paper could be produced using DNA-based self-assembly tiles \cite{Seeman1983,Rothemund2006folding,Wei2012d}, and the two-tile behaviours described in this the paper could be tested directly. Experimental and computational investigations by us in this direction are underway.

Bound and deterministic self-assembly tile models can be used as a more general language to study genotype-phenotype maps of self-assembling systems in biology, shedding light on the driving forces of Darwinian evolution and studying the balance of evolutionary forces of exploration and exploitation that act in genotype spaces \cite{Modularity,Evo,green}. This work can be extended to the non-deterministic realm, as deterministic phenotypes only occupy a relatively small fraction of genotype space, and even evolutionary advantageous phenotypes may not strictly require deterministic structural assembly if the benefits of a larger phenotype space outweigh the uncertainties of non-determinism. As this paper shows, the space of non-deterministic self-assembly exhibits a rich complexity. It is unlikely that biological evolution has left this complexity unexploited.

\bibliographystyle{unsrt}

\bibliography{PRreferences3}

\end{document}